\documentclass[conference]{IEEEtran}
\IEEEoverridecommandlockouts
\usepackage{cite}
\usepackage{amsmath,amssymb,amsfonts}
\usepackage{algorithmic}
\usepackage{graphicx}
\usepackage{textcomp}
\usepackage{xcolor}
\usepackage{url}
\def\BibTeX{{\rm B\kern-.05em{\sc i\kern-.025em b}\kern-.08em
    T\kern-.1667em\lower.7ex\hbox{E}\kern-.125emX}}
\begin{document}

\title{DeepGalaxy: Deducing the Properties of Galaxy Mergers from Images Using Deep Neural Networks\\
}


\author{\IEEEauthorblockN{Maxwell X. Cai}
\IEEEauthorblockA{\textit{SURF Corporative} \\
\textit{
Amsterdam, The Netherlands} \\
maxwell.cai@surf.nl}
\and
\IEEEauthorblockN{Jeroen B\'edorf}
\IEEEauthorblockA{\textit{minds.ai} \\
Santa Cruz, CA, USA \\
jeroen@bedorf.net}
\and
\IEEEauthorblockN{Vikram A. Saletore}
\IEEEauthorblockA{\textit{Intel Corporation} \\
Hillsboro, OR, USA \\
vikram.a.saletore@intel.com}
\and
\IEEEauthorblockN{Valeriu Codreanu}
\IEEEauthorblockA{\textit{SURF Corporative} \\
Amsterdam, The Netherlands \\
valeriu.codreanu@surf.nl}
\and
\IEEEauthorblockN{Damian Podareanu}
\IEEEauthorblockA{\textit{SURF Corporative} \\
Amsterdam, The Netherlands \\
damian.podareanu@surf.nl}
\and
\IEEEauthorblockN{Adel Chaibi}
\IEEEauthorblockA{\textit{Intel Corporation} \\
Paris, France \\
adel.chaibi@intel.com}
\and
\IEEEauthorblockN{Penny X. Qian}
\IEEEauthorblockA{\textit{Leiden University} \\
Leiden, The Netherlands \\
pennyqxr@gmail.com}
}

\maketitle

\begin{abstract}
Galaxy mergers, the dynamical process during which two galaxies collide, are among the most spectacular phenomena in the Universe. During this process, the two colliding galaxies are tidally disrupted, producing significant visual features that evolve as a function of time. These visual features contain valuable clues for deducing the physical properties of the galaxy mergers. In this work, we propose \texttt{DeepGalaxy}, a visual analysis framework trained to predict the physical properties of galaxy mergers based on their morphology. Based on an encoder-decoder architecture, \texttt{DeepGalaxy} encodes the input images to a compressed latent space $z$, and determines the similarity of images according to the latent-space distance. \texttt{DeepGalaxy} consists of a fully convolutional autoencoder (FCAE) which generates activation maps at its 3D latent-space, and a variational autoencoder (VAE) which compresses the activation maps into a 1D vector, and a classifier that generates labels from the activation maps. The backbone of the FCAE can be fully customized according to the complexity of the images. \texttt{DeepGalaxy} demonstrates excellent scaling performance on parallel machines. On the \emph{Endeavour} supercomputer, the scaling efficiency exceeds 0.93 when trained on 128 workers, and it maintains above 0.73 when trained with 512 workers. Without having to carry out expensive numerical simulations, \texttt{DeepGalaxy} makes inferences of the physical properties of galaxy mergers directly from images, and thereby achieves a speedup factor of $\sim 10^5$.
\end{abstract}


\section{Introduction}
Recent research suggests that there are up to $2 \times 10^{12}$ galaxies in the observable Universe \cite{2005ApJ...624..463G}\cite{2016ApJ...830...83C}, each of which consists of $10^{8}$ to $10^{14}$ stars. Galaxies gravitationally interact with each other during their formation and evolution histories \cite{1977egsp.conf..401T}\cite{1992ARA&A..30..705B}. As a consequence, the two interacting galaxies are shaped by the tidal field of each other, resulting to the apparent distortion in their appearance. One of the most violent interaction events is the physical collision of two galaxies, namely galaxy mergers. These spectacular events are crucial to the evolution history of galaxies, as they profoundly impact the kinematic and star formation efficiency of galaxies. While it is expected that the galaxy merger rate was higher in the early epoch of the Universe \cite{2019MNRAS.486.3702S}, galaxy mergers do still happen in the present-day Universe. For example, the Milky Way galaxy is expected to collide with the Andromeda galaxy within a timescale of $\sim 4.5$~Gyr \cite{2012ApJ...753....7S}.

The galaxy merger rate over cosmic time is one of the fundamental measures of the evolution of galaxies \cite{2011ApJ...742..103L}. The measurement of galaxy merger rates has been estimated through several approaches, including (semi-)analytical models \cite{1993MNRAS.262..627L}, on observational data \cite{2006ApJ...652..270B,2010MNRAS.401.1043D,2011ApJ...742..103L}, and cosmological simulation data \cite{2015MNRAS.449...49R}. Recently, state-of-the-art surveys have offered a wealth of observational galaxy images. Should these images be properly classified according to their morphology, a decent estimate of merger rates and galactic statistical properties can be obtained. However, the large number of images (typically of the order of $10^6$) renders it impossible for staff scientist to complete the classification themselves. For example, the \textsc{Galaxy Zoo 1} project\footnote{\url{https://www.zooniverse.org/projects/zookeeper/galaxy-zoo/}} consists of $900,000$ galaxies imaged by the Sloan Digital Sky Survey\footnote{\url{https://www.sdss.org/}}. To address this challenge, the \textsc{Galaxy Zoo} project publishes their images online, inviting citizen scientists to contribute to the classification of the galaxy images \cite{2011MNRAS.410..166L}. The citizen science approach has proven that an astrophysical problem can be translated into a pattern recognition problem, which is then efficiently tackled through the efforts of hundreds of thousands of volunteers.

In recent year, the field of image processing and pattern recognition has been boosted by the promising development of deep learning, in particular, deep convolutional neural networks (CNNs). Indeed, with the support of high-performance computing hardware, state-of-the-art CNNs such as \texttt{ResNet}\cite{he2015deep} and \texttt{EfficientNet}\cite{2019arXiv190511946T} can deliver robust results even if the images are with low signal-to-noise ratio, having obscured background, and/or distorted \cite{NIPS2012_4824,2014arXiv1409.4842S,2014arXiv1409.1556S,2015arXiv151203385H,2020ApJ...895..115F}. As such, it is feasible to automate the astronomical image classification task, which has been traditionally done by thousands of human volunteers, by using deep neural networks (DNNs). The existing classification results from human volunteers can then be used as the training datasets for building an appropriate DNN. When properly designed and trained, a DNN can deliver a speedup of many orders of magnitude compared to human classifiers, allowing astronomers to process much more data and thereby obtaining better statistics.

Roughly speaking, most galaxy image classification tasks (e.g., \cite{2018MNRAS.479..415A, 2019MNRAS.tmp.2414W, 2019PASP..131j8011R}) can be divided into the following two categories: first, it is a binary classification task to determine whether a galaxy image is a galaxy merger or a single galaxy, and second, for single galaxies, it is a multi-class classification problem to determine the galaxy type in the Hubble sequence based on its morphology. In this project, we propose \texttt{DeepGalaxy}, which is an extension to the first classification category. Instead of providing the information of whether a galaxy image is representing a galaxy merger, we aim to predict a timescale estimate at which the two merging galaxies will collide. This paper is organized as follows: The implementation of \texttt{DeepGalaxy} is presented in Section~\ref{sec:implementation}; the performance results are shown in Section~\ref{sec:results}; finally, the conclusions are summarized in Section~\ref{sec:conclusions}.

\section{Implementation}
\label{sec:implementation}
\subsection{Training Data}
\begin{figure*}
    \centering
    \includegraphics[scale=0.35]{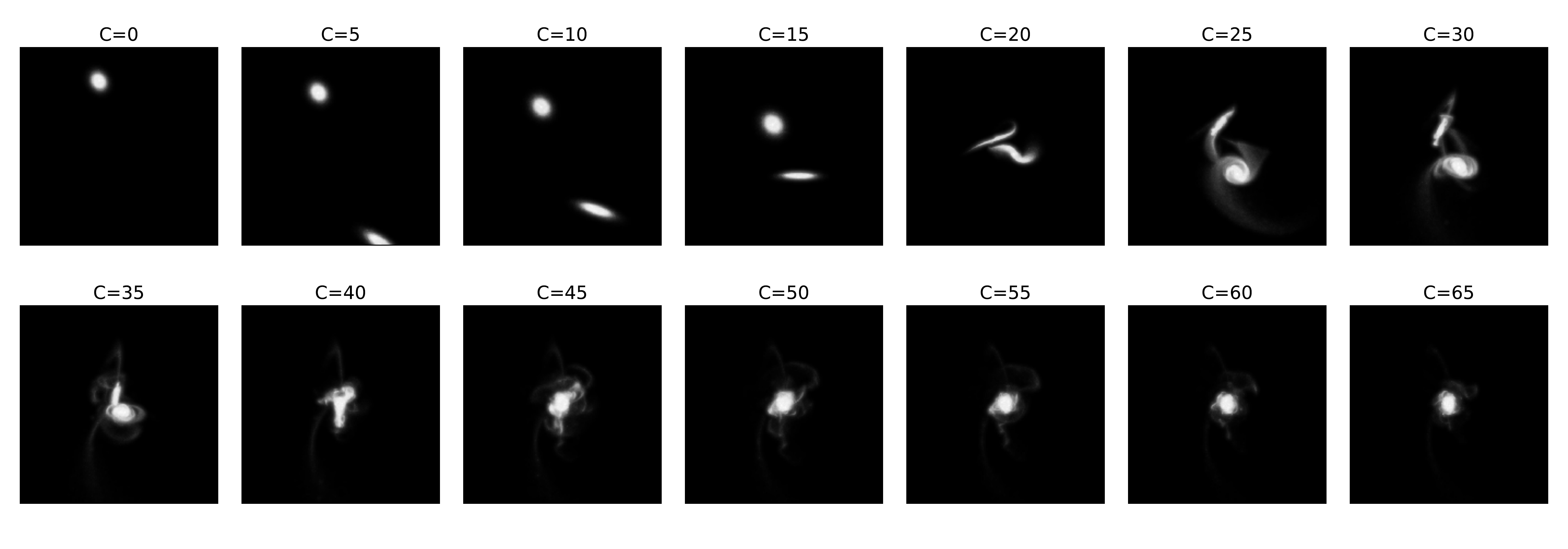}
    \caption{A gallery with a few sample images from the simulation-generated training dataset. The class labels (marked in the title of the subplots) are obtained directly from the $N$-body simulation. It indicates the timescale at which two galaxies will collide. At $C=0$, the two galaxies are still far away from each other; at $C=20$, collision is imminent; at $C=70$, the two galaxies are fully merged into an elliptical galaxy and new dynamical equilibrium has been established. }
    \label{fig:train_img_sim}
\end{figure*}

\texttt{DeepGalaxy} is a general-purpose galaxy image processing framework, so astronomers are free to choose the datasets of interests to feed into it. In our case, the training data consists of two datasets: Simulated galaxy merger images with strong labels and observation images without labels\footnote{To our best knowledge, there is no dataset of observation images with dynamical timescale labels available to the community.}. The simulated galaxy merger images and their corresponding labels are obtained through high-resolution $N$-body simulations. We use \texttt{Bonsai} \cite{Bedorf:2014:PGT:2683593.2683600, 2012JCoPh.231.2825B}, a GPU-accelerated Barnes-Hut tree \cite{1986Natur.324..446B}, code to carry out simulations of galaxy mergers with various initial conditions. The \texttt{Bonsai} simulations take into account both baryon particles and dark matter particles. The density profile of the galaxies is sampled from the Navarro-Frenk-White (NFW) profile \cite{1996ApJ...462..563N}. Each galaxy is represented by $10^7$ particles. \texttt{Bonsai} builds a Barnes-Hut tree with 19 levels according to the particle density. The simulations conserve energy rather well, up to a relative energy error of $\sim 10^{-7}$. We survey a grid of initial conditions, with the mass ratios of the two galaxies being $1/4$, $1/3$, $1/2$, $1$ and $3/2$. The two interacting galaxies are placed $\sim 20$ times their radii away from each other, such that their mutual tidal force is negligible. The two galaxies, however, are given the velocity vectors deemed for collision in the future. As they approach each other, their mutual tidal forces become increasingly stronger, which in turn shapes the morphology in this process. For each visualization time step, the galaxy merger is visualized, and the visualization is captured by virtual cameras from 14 different positions. These virtual cameras generate images of $2048 \times 2048$ pixels, large enough to resolve minute details of the merger. The dynamical timescale of the merger is encoded with 71 classes, with \texttt{class 0} representing the galaxies being the furthest away from each other (and therefore requires the longest amount of time to merge) class 70 representing the state in which the two galaxies are merged and completely virialized (i.e., reached a dynamical equilibrium state). A gallery of sample images is shown in Fig.~\ref{fig:train_img_sim}. This dataset consists of 35,784 images from 36 $N$-body simulations with different initial conditions\footnote{These 36 simulations are merely used for demo purposes. It is unlikely that they cover a full range of parameter space needed to study galaxy mergers. For actual astrophysical projects, more data is likely needed.}. The simulations are carried out using \texttt{SiMon}\footnote{\url{https://github.com/maxwelltsai/SiMon}}\cite{2017PASP..129i4503Q}, an open source simulation monitor for automating the execution of astrophysical $N$-body simulations. The dataset generated through these simulations is balanced. We randomly sample 80\% of the dataset as training data, and reserve the remaining 20\% for validation.

The dataset with observation images is derived from the \textsc{Galaxy Zoo} data. While the dataset contains labels about the galaxy morphology type, the labels are not directly usable for supervised training because they do not contain information about the dynamical timescale of the merger process. Therefore, we process this dataset with unsupervised training.

\subsection{Neural Network Architecture}
\begin{figure*}
	\centering
	\includegraphics[scale=0.25]{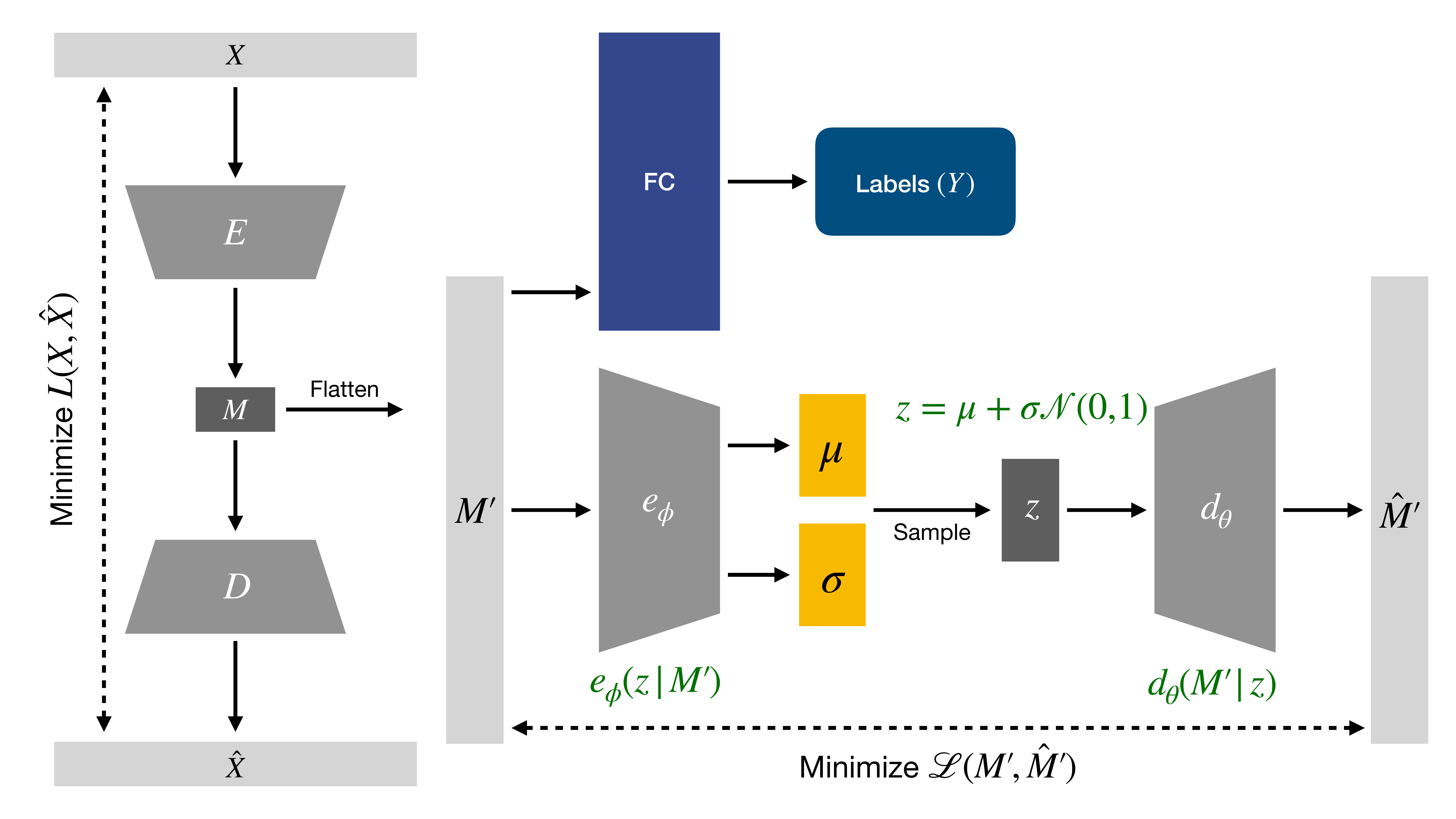}
	\caption{The architecture of \texttt{DeepGalaxy}. The DNN consists of three parts: (1) a fully convolutional autoencoder that aims to minimize the difference between the input images (denoted as $X$) and the generated images (denoted as $\hat{X}$). The encoded feature maps are denoted as $M$. (2) A fully connected network (FC) that associates the feature map $M$ with the labels $Y$. (3) A variational autoencoder that encodes the flattened feature maps $M'$ into a 1D latent vector $z$ as it reproduces $M'$. }
	\label{fig:dnn_architecture}
\end{figure*}

Having taken into account that most observation images are unlabeled, we develop a DNN architecture that supports both supervised and unsupervised training. The architecture is shown in Fig.~\ref{fig:dnn_architecture}. The DNN consists of three parts:
\begin{enumerate}
    \item Supervised learning: The input images $X$ generated from $N$-body simulations will pass through an encoder $E$, which subsequently generates feature maps $M$. The feature maps are then flattened and passed through a few fully connected layers (FC). Eventually, labels $Y$ are generated though a \texttt{softmax} activation function.
    \item Unsupervised learning of images: A fully convolutional autoencoder (FCAE) aiming to minimize the difference between $X$ and generated images $\hat{X}$. Since there is no flatten or fully-connected layer, the FCAE is invarient to image resolution.
    \item Unsupervised learning of feature maps: A variational autoencoder (VAE)\cite{2013arXiv1312.6114K} that encodes the flattened feature maps $M'$ into 1D compressed latent vectors. 
\end{enumerate}
These three parts share weights, but each of them can be trained separately. If the training data is dominated by the labeled simulated galaxy merger images, then \texttt{DeepGalaxy} becomes a standard CNN classifier, where auxiliary unlabeled images (if any) can be used as a regularization to the feature maps $M$. The backbone of the convolutional encoder $E$ and convolutional decoder $D$ can either be a simple network or a state-of-the-art one with residual blocks (e.g., \texttt{ResNet} and \texttt{EfficientNet}). The choice of the backbone depends on the complexity of the input data. On the other hand, if the data is dominated by unlabeled observational data, \texttt{DeepGalaxy} can be used as an unsupervised clustering algorithm, and the labeled images can be considered as a regularization of $M$. If comparing the image similarity is the main interest, then \texttt{DeepGalaxy} can be considered as an image search framework, where images are compared according to their distances in the latent space $z$. By training the VAE with the high-level representation $M$ instead of the pixel-level raw images $X$, the VAE has access to the global features of $X$, making it more robust. The encoded latent vector $z$ is a 1D representation with a small size (typically with the length of 16 or 32). As such, comparing images via $z$ is computationally much more affordable than comparing directly at the pixel level.

The current implementation of \texttt{DeepGalaxy} is based on TensorFlow. 

\subsection{Parallelization}
As mentioned above, the backbone of $E$ and $D$ can be chosen according to the actual complexity of the images. In the scenario where processing high-resolution (observational) images is the main interest, it is usually necessary to use a large and deeper convolutional network as the backbone. Notably, the newly developed CNN \texttt{EfficientNet} has fewer parameters, but higher accuracy than \texttt{ResNet50}. But even so, when using the largest variant of \texttt{EfficientNet},  \texttt{EfficientNetB7}, to process images with a resolution of $(2048 \times 2048)$ pixels, requires roughly 66 million parameters with a memory footprint of 150 GB (when using a batch size of 1). Consequently, in this regime, parallelization becomes particularly important, even required. We implement the parallelization using \texttt{Horovod}, a data parallelism framework built on the top of popular deep learning frameworks (such as TensorFlow and PyTorch) and high-performance collective communication frameworks (such as MPI and NCCL). As such, \texttt{DeepGalaxy} can be easily scaled up in a multi-node, multi-worker high-performance computing environment.

\subsection{Hardware Environment}
We carry out the training and benchmark of \texttt{DeepGalaxy} on the \emph{Endeavour} HPC Supercomputer hosted at Intel Corporation. Since the memory footprint required for the training is usually larger than the available memory of the latest GPU, the training is carried out on the CPU. We use up to 256 nodes on \emph{Endeavour}, each node with two Intel(R) Xeon (R) Scalable Processor 8260 CPU and 192 GB of memory. The compute nodes are interconnected with the Intel(R) Omni-Path Architecture.

\section{Results}
\label{sec:results}

\subsection{Classification accuracy}
\begin{figure}
	\centering
	\includegraphics[scale=0.42]{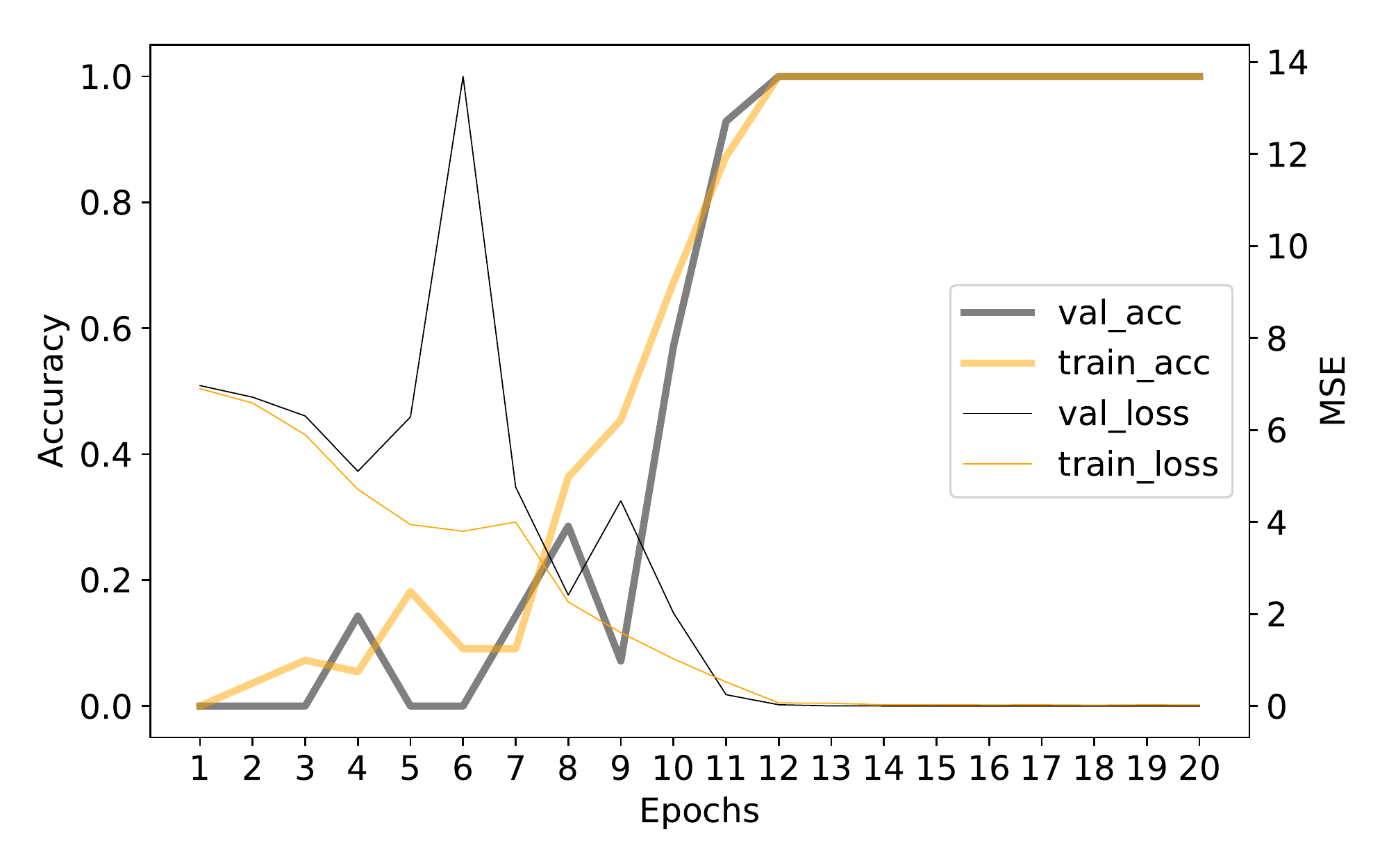}
	\caption{The loss and accuracy of the training dataset and the validation dataset as a function of epochs. The loss is measured with mean squared error (MSE). The image resolution is $512 \times 512$ pixels. The CNN backbone is \texttt{EfficientNetB4}, initialized with the \texttt{ImageNet} weights. The training is carried out on 256 nodes, each with 2 workers (thus 512 workers in total). The network is initialized with \texttt{ImageNet} weights. We use a momentum of 0.9 in the batch normalization layers. }
	\label{fig:training}
\end{figure}

For the convolutional classifier, we measure the classification accuracy on an \texttt{EfficientNetB4} network, which consists of roughly 18 million parameters (comparable to \texttt{ResNet50}, which has about 23 million parameters). The network is initialized using \texttt{ImageNet}, and is subsequently trained with simulated galaxy merger images with the size of $(512 \times 512)$ pixels \footnote{While the original resolution of images generated from the simulations is $(2048 \times 2048)$ pixels, we down-sampled these images to $(512 \times 512)$ pixels to quicken the training process.}. We use \texttt{Optuna}\footnote{\url{https://optuna.org/}}, a hyper-parameter optimization tool to obtain the optimal learning rate. With the \texttt{AdaDelta} optimizer and a learning rate of 8.0, the network converges within about 12 epochs, as shown in Fig.~\ref{fig:training}. In comparison, without the \texttt{ImageNet} weights initialization, it will take 40 epochs of training to converge. The local batch size is 8 and with 512 workers, the global batch size is 4096. This setting results in a memory footprint of 16 GB per worker. With 256 computing nodes on the \emph{Endeavour} supercomputer, convergence is reached within 15 minutes.

\begin{figure}
	\centering
	\includegraphics[scale=0.42]{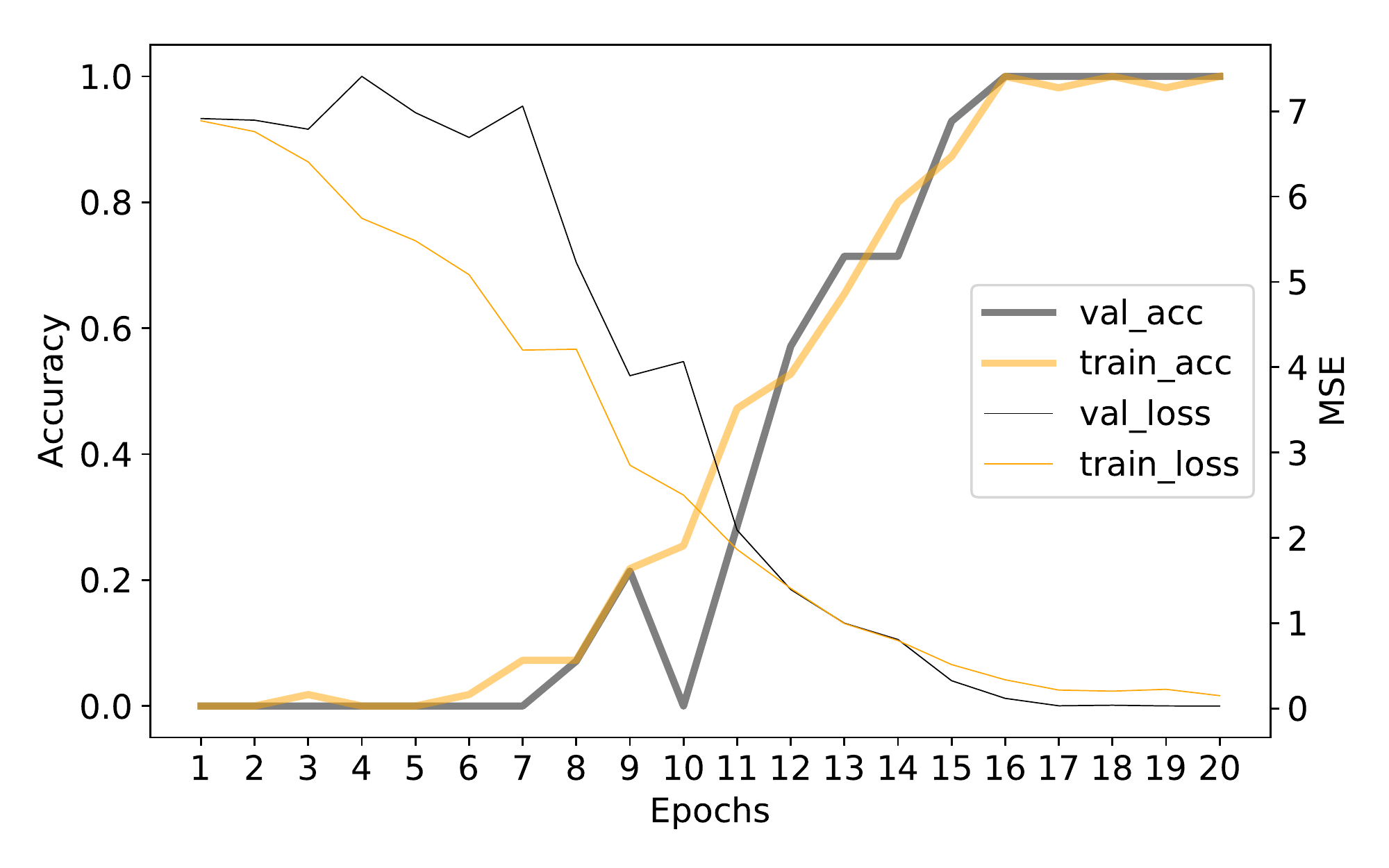}
	\caption{Same as Fig.~\ref{fig:training}, but for the images of $(1024 \times 1024)$ pixels.}
	\label{fig:training_1024}
\end{figure}

In addition, to study the how the image resolution affect the convergence, we train another \texttt{EfficientNetB4} network on images with $(1024 \times 1024)$ pixels. Using the same hardware configuration, with a local batch size of 8 (global batch size 4096), convergence is reached within 16 epochs, as shown in Fig.~\ref{fig:training_1024}. It has a memory footprint of 43 GB per worker, which is too large to fit into the GPU memory. With 256 computing nodes on the \emph{Endeavour} supercomputer, convergence is reached within 15 minutes.

\subsection{Unsupervised Learning Performance}
\begin{figure}
    \centering
    \includegraphics[scale=0.33]{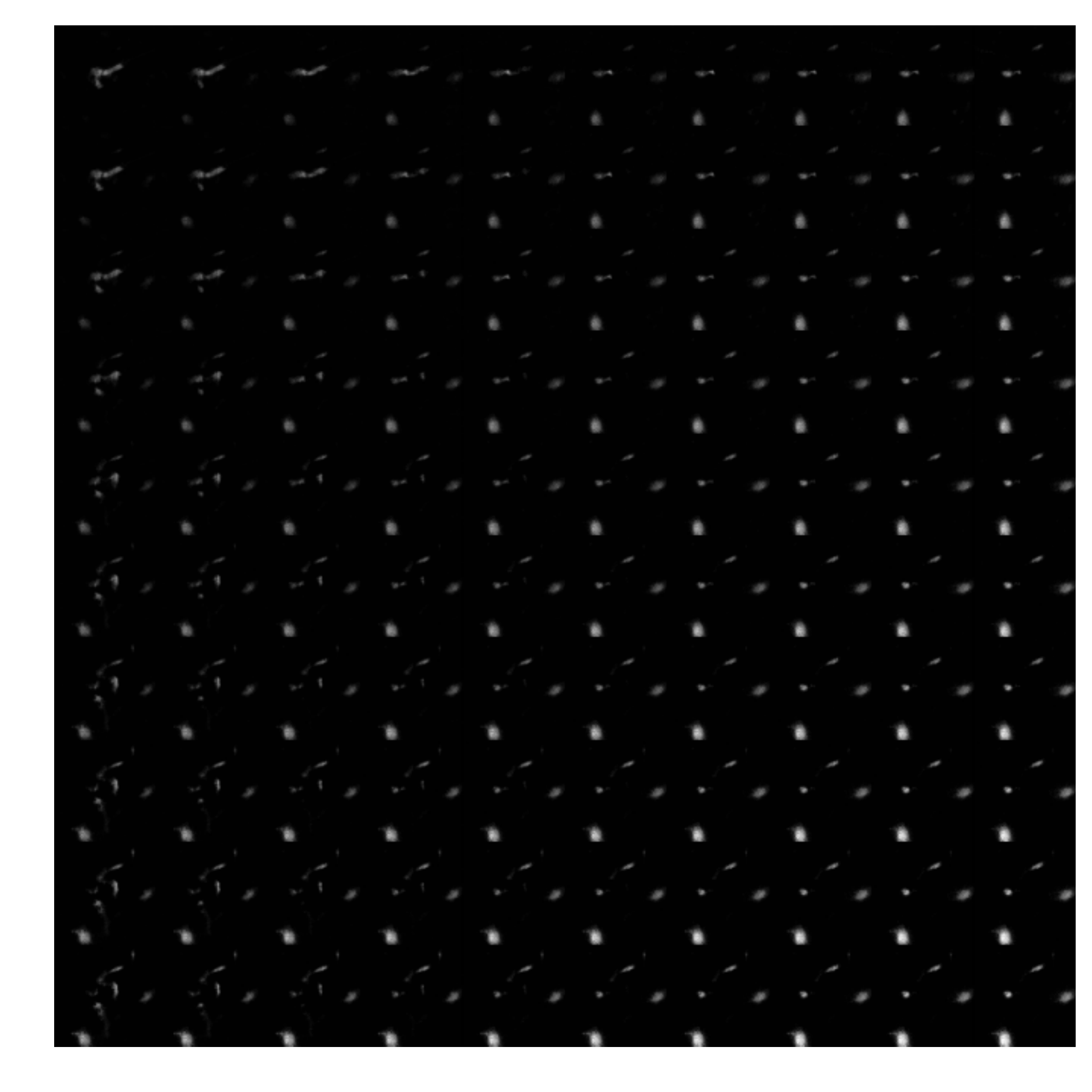}
    \caption{Galaxy merger images generated by the VAE when interpolating the latent vector $z$.}
    \label{fig:vae_latent_images}
\end{figure}

\begin{figure*}
    \centering
    \includegraphics[scale=0.6]{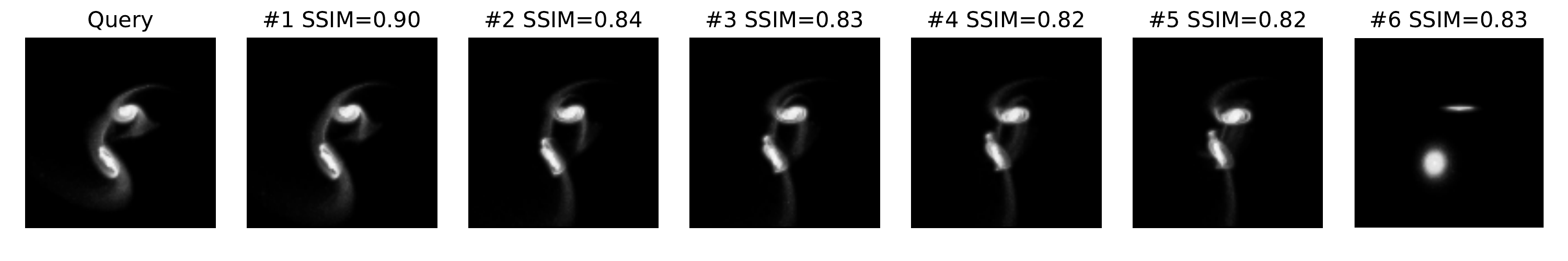}
    \caption{The latent variable model (consisting of a convolutional feature encoder, a variational autoencoder, and a convolutional feature decoder) is able to generate a few suggestions based on the query images (plotted on the left). The suggestions are sorted by descending similarity, where 1 indicates a high-degree of similarity and 6 indicates a low-to-moderate degree of similarity. The similarity of images is quantified with the structural similarity index measure (SSIM), in which a score of 1.0 means identical, and a score of 0 means completely different.}
    \label{fig:vae_suggestions}
\end{figure*}

As shown in Fig.~\ref{fig:vae_latent_images}, the VAE manages to learn an embedding in its latent space $z$. As such, when interpolating $z$, a grid of smoothly varying images is generated. Therefore, image similarity can be measured by calculating $||z_1 - z_2||$, where $z_1$ is the latent vector of a query image and $z_2$ is the latent vector of a training image. By getting the $k$-nearest neighbor (kNN) vectors around the query vector $z_1$, the latent model is essentially suggesting $k$ similar images, and therefore the physical properties of the query images can be inferred from that of a similar figure. However, this algorithm requires defining a number of neighbors, which could be a limitation. In order to get fully unsupervised recommendations for an arbitrary query, we continue the previous algorithm and build the kNN graph based on distances between samples. We then calculate the discrete Laplacian, obtaining a matrix which measures to what extent the graph differs at one vertex, sample, to nearby vertices, other samples. This allows us to explore the connected components and the spectrum of the graph. Finally we perform a spectral clustering step in order to obtain the final recommendations. The advantage of exploring the graph structure in this way is that the initial choice for $k$ becomes less important. When running this query thought the synthetic dataset, we validate the recommendation quality against the structural similarity index measure (SSIM). Fig.~\ref{fig:vae_suggestions} demonstrates the effectiveness of this approach. The recommendations generated by the VAE have yielded a high degree of consistency with the SSIM score. It is interesting to note that, although the SSIM score of the last recommendation (\#6) has a slightly higher SSIM score than that of recommendation \#5, what the VAE has recommended appears to be more visually convincing. When querying all samples from the dataset, we obtain an average SSIM score of 0.9.

\subsection{Scaling Performance}
\begin{figure}
    \centering
    \includegraphics[scale=0.45]{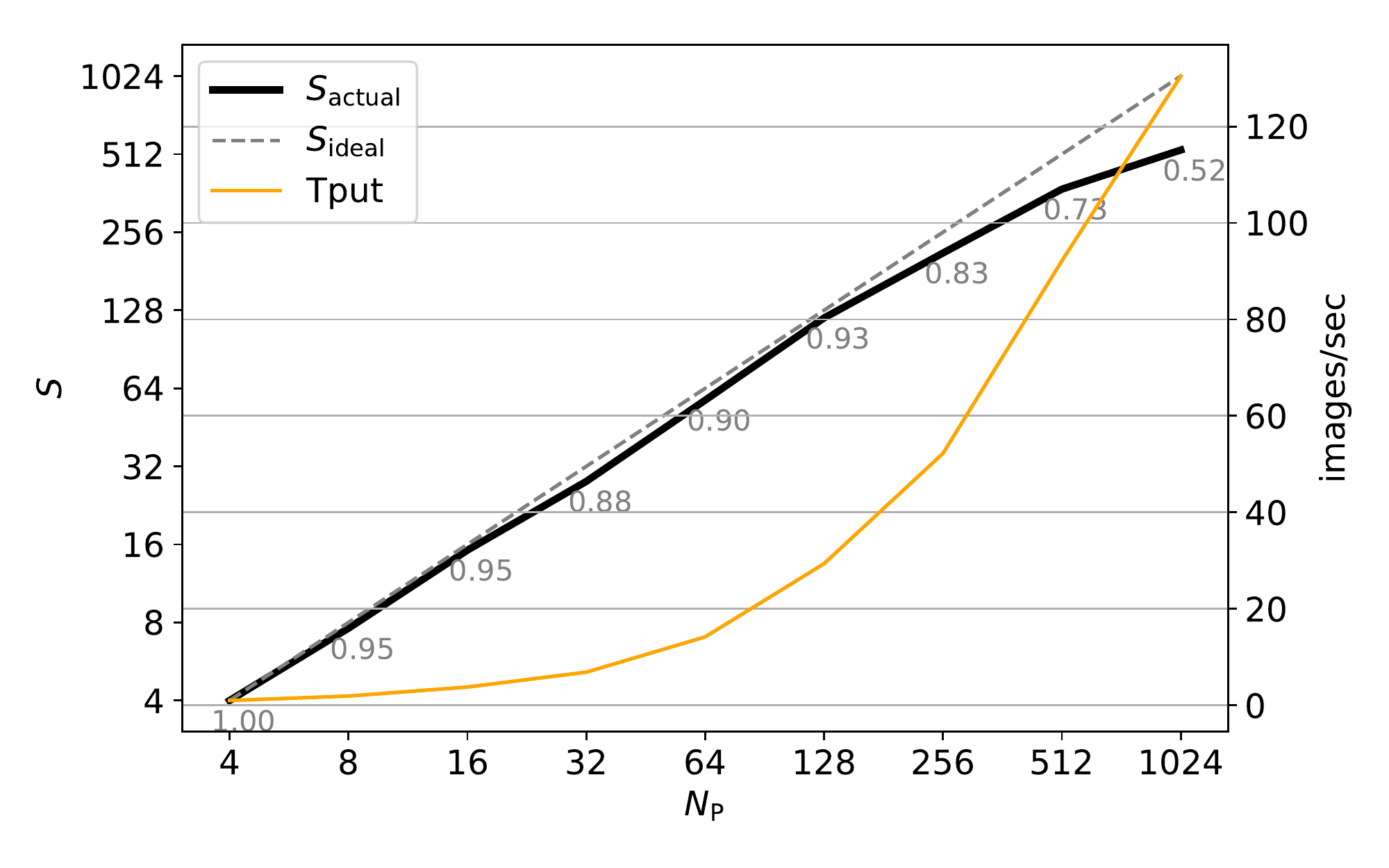}
    \caption{The scaling performance of \texttt{DeepGalaxy} as a function of the number of workers $N_{\rm P}$. Left $y$-axis: The speed up ratio $S$ quantifies the throughput ratio between $N_{\rm P}$ nodes and one node. In the ideal case, $S = N_{\rm P}$ (dashed diagonal line), but the actual measurement (indicated with thick black curve) is smaller than $N_{\rm P}$ due to the gradient communication and other overheads. Right $y$-axis: The throughput (Tput) is shown with the orange curve, which quantifies the number of images trained per second by the whole system. Grey numbers along the thick black curve: The scaling efficiency as a function of $N_{\rm P}$ (normalized to $N_{\rm P} = 4$), which is a value with the range of $[0,1]$. A scaling efficiency of 1 indicates perfect scaling. This benchmark is carried out with the $(1024 \times 1024)$ pixels images with the \texttt{EfficientNetB4} network on the \emph{Endeavour} supercomputer. For the case of $N_{\rm P} = 512$, the training accuracy/loss and validation accuracy/loss are shown in Fig.~\ref{fig:training}.}
    \label{fig:scaling}
\end{figure}

We perform the training on different number of nodes on the \emph{Endeavour} cluster, and thereby obtaining a scaling performance. As shown in Fig.~\ref{fig:scaling}, \texttt{DeepGalaxy} is configured with an \texttt{EfficientNetB4} backbone network, and is subsequently trained on the $(1024 \times 1024)$ pixel images. With the number of workers $N_{\rm P}$ ranging from 4 to 1024 (corresponding to 1 to 256 nodes, each node with 4 workers), the throughput scales nearly linearly as a function of $N_{\rm P}$ for $N_{\rm P} \leq 128$. The scaling efficiency for $N_{\rm P} = 128$ exceeds 0.93. The communication overhead starts to pick up in the case of $N_{\rm P} > 128$. Nevertheless, the scaling efficiency remains significant even in the case of $N_{\rm P} = 512$, hitting a value of 0.73.  This figure, therefore, demonstrates that \texttt{DeepGalaxy} is able to handle very large and complex datasets when scaling up on massively parallel computers.

\subsection{Inference Speed-Up}
When using the \texttt{FP32} precision, it takes the training model $\sim 2$ seconds to perform the inference classification operation.  In contrast, it would take approximate 1-2 days with the traditional methods to carry out numerical simulations of the galaxy merger and make the prediction of its properties (depending on the initial conditions and the resolution of the simulation). As such, a speed up factor of $10^4 - 10^5$ is achieved when using \texttt{DeepGalaxy} to replace traditional numerical simulations. Greater speed-up ratios can be achieved when using reduce-precision inference. Such a speed up factor over simulation is extremely promising for tackling the very large amount of galaxy merger data, as this allows astrophysicists to quickly obtain the statistical properties of a large number of samples.

\section{Conclusions}
\label{sec:conclusions}
Big Data is a constant challenge in modern astrophysical research. With terabytes of images generated by telescopes every night, astronomers usually would have to depend on the help of thousands of volunteers to process the data. In recent years, image processing and pattern recognition have significantly advanced thanks to the promising development of deep neural networks (DNNs). With state-of-the-art DNN architectures and high-performance computing technologies, it is increasing feasible to make use of DNNs for constructing highly efficient and robust astrophysical imagery data processing pipelines.

In this project, we develop a suite of DNNs, namely \texttt{DeepGalaxy}, to process classification and unsupervised analysis (e.g., morphology matching) of galaxy images. We present an example of using \texttt{DeepGalaxy} to predict the dynamical timescales of major galaxy mergers from their images. Constructing the intrinsic properties or initial conditions of a physical system based on observations is a typical inverse problem, which, traditionally, requires carrying out extensive astrophysical $N$-body simulations. \texttt{DeepGalaxy}\footnote{\url{https://github.com/maxwelltsai/DeepGalaxy}} aims to bypass the need of carrying out simulations. Trained with both simulation data and observational data, \texttt{DeepGalaxy} aims to extract the information directly from the morphology of galaxies. It can be used as a convolutional classifier, an unsupervised clustering algorithm, and an image pairing algorithm. Without having to carry out numerical simulations, this approach achieves a speedup of $10^4 - 10^5$.

As a general-purpose galaxy image processing framework, \texttt{DeepGalaxy} is designed for a wide range of galaxy image datasets. Depending on the complexity of the input image data, \texttt{DeepGalaxy} allows the users to freely choose the architecture of the convolutional backbone as needed. Furthermore, the training can be scaled up efficiently on massively parallel machines (tested on up to 256 nodes, 1024 workers), either on the CPUs or on the GPUs. On the \emph{Endeavour} supercomputer, we obtain a scaling efficiency of 0.93 when trained with 128 workers, and an efficiency of 0.73 when trained with 512 workers.


\section*{Acknowledgment}
We thank the anonymous referees for their constructive comments, which helped to improve the paper. We thank Malavika Vasist, Bruno Alves, Simon Portegies Zwart, Reinout van Weeren, and Hans Pabst for the numerous insightful discussions that we have had so far.


\bibliographystyle{ieeetr}
\bibliography{refs.bib}

\end{document}